\documentclass[12pt,aps,pra,preprint]{revtex4}
\usepackage{graphicx}
\usepackage{amsmath,amssymb,amsfonts}
\usepackage{natbib}

\newcommand{\bd}{\begin{displaymath}}
\newcommand{\ed}{\end{displaymath}}
\newcommand{\be}{\begin{equation}}
\newcommand{\ee}{\end{equation}}
\newcommand{\bs}{\begin{subequations}}
\newcommand{\es}{\end{subequations}}
\newcommand{\ba}{\begin{eqnarray}}
\newcommand{\ea}{\end{eqnarray}}

\begin{document}

\title{Aspects of nonlocality from a quantum trajectory perspective:\\
A WKB approach to Bohmian mechanics}

\author{A. S. Sanz}
\email{asanz@imaff.cfmac.csic.es}

\author{S. Miret-Art\'es}
\email{s.miret@imaff.cfmac.csic.es}

\affiliation{Instituto de Matem\'aticas y F\'{\i}sica Fundamental\\
Consejo Superior de Investigaciones Cient\'{\i}ficas\\
Serrano 123, 28006 Madrid, Spain}

\date{\today}

\begin{abstract}
Nonlocality is a property of paramount importance both conceptually and
computationally exhibited by quantum systems, which has no classical
counterpart.
Conceptually, it is important because it implies that the evolving
system has information on what happens at any space point and time.
Computationally, because such a knowledge makes any calculation
intractable as the number of degrees of freedom involved increases
beyond a few of them.
Bohmian mechanics, with its trajectory-based formalism in real
configuration space, can help to better understand nonlocality.
A detailed analysis of how nonlocal information is transmitted to
quantum trajectories in simple systems (free particle and harmonic
oscillator) turns out to be very interesting when compared to analogous
systems in classical mechanics.
\end{abstract}

\maketitle


\section{Introduction}

Since the early steps of quantum theory there has always been much
interest in establishing a connection between this theory and classical
mechanics.
A lot of work has been devoted to derive quantization rules based on
the correspondence principle \cite{paul1}, to propose and improve
semiclassical approximations \cite{grossman,miller,eli,sanz4}, or to
develop quantum-classical studies by using the Liouvillian formulation
of quantum mechanics \cite{paul21,paul22,paul31,paul32} or some sort
of approximations where quantum-classical mixed dynamics are involved
\cite{alberto}.
However, there is an alternative pathway consisting of exploiting
the properties offered by the so-called quantum Hamilton-Jacobi
equation, as done by Leacock and Padgett \cite{leacock} in 1983.
This equation can be derived from the time-dependent
Schr\"odinger equation by means of a simple transformation which allows
to express the wave function in terms of a complex phase (which also
contains all the information that carries the wave function).
The formulation based on the quantum Hamilton-Jacobi equation was
formerly aimed to obtain bound-state energy levels of quantum systems
with no need to solve the equation of motion involved in the
calculation of the system wave function.

Leacock and Padgett formulation leads straightforwardly to quantum
trajectory approaches, which have also been an alternative to
understand the relationship between quantum and classical mechanics.
Bohmian mechanics, which was proposed before, is also one of these
approaches.
Nonetheless, it presents an inconvenient when dealing with (energy)
eigenstates.
These states are associated with zero velocity fields, and therefore
particles remain steady when described by them \cite{holland}.
To solve this problem, Floyd \cite{floyd1,floyd2} and Faraggi and
Matone \cite{faraggi1,faraggi2} developed quantum Hamilton-Jacobi
like formulations starting from (real) {\it bipolar ansatzs}, though
they did not claim full equivalence with standard quantum mechanics
regarding their predictions.
This problem was also treated by John \cite{john} by means of a theory
based on {\it complex quantum trajectories} derived from the ansatz
used by Leacock and Padgett, and that was applied to some simple
analytical cases, such as the harmonic oscillator and the step barrier.
This trajectory-based approach has been used recently
\cite{wyatt1,wyatt2,tannor,sanz0} to propose different numerical
algorithms that allow to obtain transition probabilities and
bound-state energies.

When quantum trajectory-based approaches are used, a full
{\it localization} underlies the particle dynamics since we are
dealing with well-defined trajectories in (either real or complex)
configuration space.
Then, one might have the wrong impression that nonlocal effects are
somewhat washed out.
Remember that nonlocality is present in so remarkable phenomena such as
entanglement, which is the cornerstone of quantum information theory and
quantum computation, for instance.
Hence nonlocality becomes an important issue when working with quantum
trajectories (mainly when proposing alternative methods to solve
quantum problems by means of them), because not only it does not
disappear at all, but plays a key role in the particular topology
displayed by such trajectories.
Independently of connotations coming from entanglement and quantum
information theory, nonlocality is present even when considering  the
time evolution of a single particle.
This can be nicely seen by means of Bohmian mechanics, where quantum
trajectories are defined in the (real) configuration space where
experiments are carried out.
In order to reproduce experimental outcomes, one needs to collect many
particles which (we assume) describe certain trajectories within the
Bohmian approach.
All these particles are independent (i.e., there is no physical
interaction among them) and reach the detector (whatever it is) one
by one.
Despite their independence, all of them are ``guided'' by identical
initial wave functions (or wave fields); the corresponding trajectories
are then obtained from the time evolution of the wave field, and the
outcomes from the asymptotic limits in space and time of $\rho(x,t) =
|\Psi(x,t)|^2$ (we can consider one dimension in this description
without loss of generality).
The effects of $\Psi$ on the particle motion can be then considered as
a manifestation of how nonlocality arises in quantum mechanics.

In this Letter a discussion on nonlocality is carried out in terms of
the WKB-like formulation of Bohmian mechanics (WKB-BM), proposed by
Sanz {\it et al.}\ \cite{sanz4,sanz1,sanz10} to understand the
quantum-classical correspondence.
This correspondence has been recently considered in the study of
different problems appearing in realistic atom-surface scattering
problems \cite{sanz2,sanz21,sanz22}.


\section{WKB formulation of Bohmian mechanics}

The starting point of the WKB-BM approach consists of using the
general ansatz \cite{leacock,john,sanz1}
\begin{equation}
 \Psi (x,t) = e^{i\bar{S}(x,t)/\hbar} ,
 \label{eq1}
\end{equation}
where $\bar{S}$ is a complex (phase) function.
This simple relation allows to express the time-dependent
Schr\"odigner equation as a quantum Hamilton-Jacobi equation,
\begin{equation}
 \frac{\partial \bar{S}}{\partial t}
  + \frac{\left( \nabla \bar{S} \right)^2}{2 m}
  + V - \frac{i\hbar}{2m} \nabla^2 \bar{S} = 0 .
 \label{eq2}
\end{equation}
Equivalently, applying the $\nabla$ operator, this equation can be
expressed in its hydrodynamical form,
\begin{equation}
 \frac{d \bar{v}}{dt} = - \frac{\nabla V}{m}
  + \frac{i\hbar}{2m} \ \! \nabla^2 \bar{v} ,
 \label{eq3}
\end{equation}
where $\bar{v} = \nabla\bar{S}/m$ is a complex velocity field and
$d/dt = \partial/\partial t + \bar{v}\nabla$ is the associated
Lagrangian operator.
This equation is the starting point of different numerical algorithms
proposed in the literature \cite{tannor,wyatt3} to deal with a large
variety of quantum problems, particularly high-dimensional ones, where
standard quantum mechanics is difficult to apply.

After Eq.~(\ref{eq2}), the WKB-BM approach considers an expansion
of $\bar{S}$ as a power series of $\hbar/i$,
\begin{equation}
 \bar{S}(x,t) = \sum_{n = 0}^\infty
  \left( \frac{\hbar}{i} \right)^n \bar{S}_n (x,t) ,
 \label{eq4}
\end{equation}
where the functions $\bar{S}_n$ are real (as seen below, they satisfy
a hierarchy of real coupled equations).
We are interested in knowing how and when nonlocality appears, and
therefore whether a motion can or cannot be regarded as classical.
Therefore, the functions $\bar{S}_n$ have to obey the usual conditions
of the standard WKB approximation \cite{gottfried}.
Substituting (\ref{eq4}) into Eq.~(\ref{eq2}), the latter can
be conveniently decoupled in powers of $\hbar/i$ to yield
\begin{equation}
 \frac{\partial \bar{S}_0}{\partial t}
  + \frac{( \nabla \bar{S}_0 )^2}{2m} + V = 0
 \label{eq5}
\end{equation}
for the zeroth order (i.e., the classical Hamilton-Jacobi equation,
with $\bar{S}_0$ being the classical action), and
\begin{equation}
 \frac{\partial \bar{S}_n}{\partial t} +
  \frac{1}{2 m} \ \! \sum_{k = 0}^n
  \nabla \bar{S}_k \nabla \bar{S}_{n-k} +
  \frac{1}{2 m} \ \! \nabla^2 \bar{S}_{n-1} = 0
 \label{eq6}
\end{equation}
for higher orders, $n \geq 1$.
Eq.~(\ref{eq6}) together with Eq.~(\ref{eq5}) provides
an iterative scheme to compute higher orders in the expansion
of $\bar{S}$.
In particular, when the series is truncated at $n=1$ one gets the
well-known WKB approximation.
This scheme is not based on the propagation of individual trajectories,
but on the propagation of functions displaying the  global behavior of
ensembles of particles.
In this sense, the $\bar{S}_n$ functions can be considered as fields.

Once every $\bar{S}_n$ is known, the series (\ref{eq4}) will also
be known.
One can then compute the real fields $R$ and $S$ that define the
Bohmian (polar) ansatz
\be
 \Psi (x,t) = R(x,t) e^{i S(x,t)/\hbar}
 \label{eq7}
\ee
in terms of the $\bar{S}_n$ functions by identifying Eqs.~(\ref{eq1})
and (\ref{eq7}) to give
\bs
\begin{eqnarray}
 R & = & \exp \left[ \sum_{n=0}^\infty (-1)^n \hbar^{2n} \ \!
   \bar{S}_{2n+1} \right] ,
 \label{eq8}
  \\ & & \nonumber \\
 S & = & \sum_{n=0}^\infty (-1)^n \hbar^{2n} \ \! \bar{S}_{2n} .
 \label{eq9}
\end{eqnarray}
\es
We would like to stress that although $S$ is given in terms of even
powers of $\hbar$, eventually one could observe constant or
time-dependent terms depending linearly on $\hbar$ (see below).
Note also that the wave function is always well defined except an
overall phase that does not depend on the coordinate space, which
is not taken into account in~(\ref{eq1}).
By substituting Eq.~(\ref{eq9}) into the Bohmian velocity field,
\be
 \dot{x} = v = \frac{\nabla S}{m} ,
 \label{eq10}
\ee
we reach
\be
 \dot{x} = \dot{x}^{({\rm cl})} + \frac{1}{m} \sum_{n=1}^\infty
    (-1)^n \hbar^{2n} \ \! \nabla \bar{S}_{2n} ,
 \label{eq11}
\ee
where $\dot{x}^{({\rm cl})} = \nabla \bar{S}_0/m$ is the classical
law of motion.
From Eq.~(\ref{eq11}) one can therefore interpret quantum
trajectories as classical trajectories ``dressed'' with a series of
terms coming from quantum interference.
This result makes apparent the important difference between both types
of trajectories, classical and Bohmian.
Within the WKB-BM approach, nonlocal effects are contained in the
second term of the r.h.s.\ of Eq.~(\ref{eq11}) and start appearing at
very small times.


\section{Applications of the WKB-BM approach}

In the light of the WKB-BM approach the effects of the nonlocality
are nicely illustrated by means of the following well-known examples.
First, we consider the case of the free propagation of a Gaussian wave
packet,
\be
 \Psi (x,t) = \left( \frac{1}{2\pi\tilde{\sigma}_t^2}\right)^{1/4}
  e^{-(x - v_0 t)^2/4\tilde{\sigma}_t\sigma_0
   + i p_0 (x - v_0 t)/\hbar + i E t/\hbar} ,
 \label{eq12}
\ee
where the time complex spreading is given by $\tilde{\sigma}_t = \sigma_0
(1 + i\hbar t/2m\sigma_0^2)$, the initial velocity by $v_0 = p_0/m =
\langle \hat{p}/m \rangle$, and the initial energy by
$E=p_0^2/m = \langle \hat{H} \rangle$.
The corresponding (Bohmian) action is then
%
\be
 S (x,t) = - \frac{\hbar}{2} \ \! \arctan \left(\frac{\hbar t}
  {2m\sigma_0^2} \right) + E t + p_0 x
   + \frac{\hbar^2 t}{8m\sigma_0^2\sigma_t^2} \ \! (x - v_0 t)^2 ,
 \label{eq13}
\ee
with $\sigma_t = \sigma_0 [1 + (\hbar t/2m\sigma_0^2)^2]^{1/2}$ being
the time-dependent real spreading.
For $\hbar$ small (or, in general, $\hbar t/2m\sigma_0^2$ small),
note that $S$ can be expressed, effectively, as a series of even
powers of $\hbar$ when the arctangent function and the spreading
of the wave packet are expanded in a Taylor expansion.

Introducing (\ref{eq13}) into Eq.~(\ref{eq10}), we obtain the exact
expression for the quantum trajectories,
\be
 x(t) = v_0 t + \frac{\sigma_t}{\sigma_0} \ \! x_0 ,
 \label{eq14}
\ee
where the initial condition is $x(0) = x_0$.
If we assume $\hbar t/2m\sigma_0^2 < 1$, (\ref{eq14}) can be
expressed as
%
\be
 x(t) = x_0 + v_0 t + \sum_{n=1}^\infty (-1)^{(n-1)}
  \ \! \frac{(2n-3)!!}{2^n \cdot n!} \ \!
   \left(\frac{\hbar t}{2m\sigma_0^2}\right)^{2n} \ \! x_0 .
 \label{eq15}
\ee
As expected, the two first terms in the r.h.s.\ of this expression are
exactly the same that one would expect from a classical trajectory;
note that the first term arises precisely when there is no spreading.
However, the presence of the $\hbar$-dependent (third) term makes
that a divergence with respect to the classical motion be observable:
it leads to a hyperbolic spreading of the trajectories.
This is a very remarkable effect where the nonlocality, as defined
above, plays an important role; in order to avoid crossing among
trajectories, those with the outmost initial conditions (with respect
to $x_0 = 0$) will spread at a faster rate than the innermost ones.
This information is contained in the spreading ratio $\sigma_t/\sigma_0$,
which provides {\it global} information (i.e., it transmits the
information carried by the wave function) on how the full velocity
field has to be at any time.
In other words, in this simple example the second term of
Eq.~(\ref{eq14}) is responsible for nonlocal effects.
Nevertheless, note that for relatively small $\hbar$ (or, equivalently,
very short timescales, wide wave packets, and/or massive particles), the
quantum trajectories will basically follow a similar behavior as
classical particles, this explaining the good agreement between quantum
mechanics and semiclassical approaches in this case.

\begin{figure}
 \includegraphics[width=7cm]{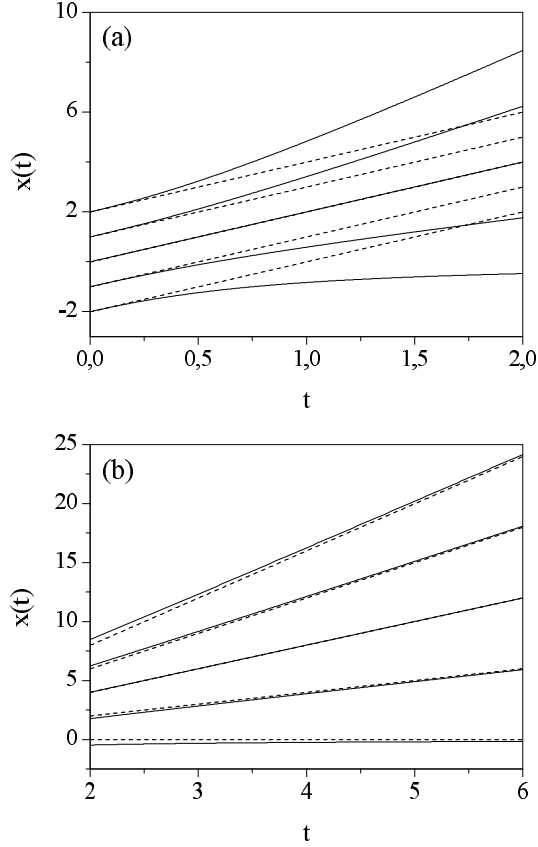}
 \caption{\label{fig1}
  Bohmian trajectories (solid lines) corresponding to a free Gaussian
  wave packet and obtained by means of Eq.~(\ref{eq14}) at:
  (a) short timescales (region around $\hbar t/2m\sigma_0^2 \lesssim 1$)
  and (b) asymptotic times ($\hbar t/2m\sigma_0^2 \gg 1$).
  To compare, the corresponding classical trajectories have also been
  represented with dashed lines.
  In part (b), dashed lines indicate the classical-like asymptotes
  described by Eq.~(\ref{eq16}).}
\end{figure}

The general trend described above is illustrated in Fig.~1, where
a sample of Bohmian and classical trajectories with equal initial
conditions are shown.
Observe that after a very short time, Bohmian trajectories start
displaying a completely different behavior than their classical
counterparts.
Moreover, note also that at $\hbar t/2m\sigma_0^2 \sim 1$ the series
given by Eq.~(\ref{eq15}) breaks down and can no longer be used to
describe the long-time behavior of the Bohmian trajectories.
It is then interesting to consider the asymptotic limit of
Eq.~(\ref{eq14}), given by
\be
 x(t) \approx
  \left( v_0 + \frac{\hbar x_0}{2m\sigma_0^2} \right) t .
 \label{eq16}
\ee
This result shows us that asymptotically (within the so-called
Fraunhofer regime of motion \cite{sanz1,sanz10,sanz7,sanz8}) quantum
trajectories (as classical trajectories) describe a uniform rectilinear
motion.
However, the corresponding quantum constant velocity given by
$v'_0 = v_0 + \hbar x_0/2m\sigma_0^2$ contains a residual term coming
from the nonlocality (the velocity of each particle depends on its
corresponding initial position). As discussed elsewhere
\cite{sanz1,sanz10,sanz2,sanz21,sanz22,sanz7,sanz8}, the nonlocal and
context-dependence
information transmitted to the quantum trajectories is through the
quantum potential, which is related to the curvature of the wave
function and therefore is responsible for the residual contribution
of the constant velocity.

Next example is the harmonic oscillator, which is also worth
discussing.
In this case, the time-evolution of a Gaussian wave packet centered
around $x=a$ is described by
%
\be
 \Psi (x,t) = \left( \frac{1}{2\pi\sigma_0^2} \right)^{1/4}
  e^{-(x - a \cos \omega t)^2 / 4\sigma_0^2 - i \omega t/2
   - i m\omega (4 x a \sin \omega t - a^2\sin 2\omega t) / 4\hbar} ,
 \label{eq17}
\ee
where $\sigma_0^2 = \hbar/2m\omega$. Here, the quantum action is
\be
 S (x,t) = - \frac{1}{2} \ \! \hbar\omega t
  - \frac{m\omega}{4} \ \! (4xa\sin \omega t - a^2\sin 2\omega t) ,
 \label{eq18}
\ee
and therefore the quantum trajectory is given by
\be
 x(t) = (x_0 - a) + a \cos \omega t ,
 \label{eq19}
\ee
which satisfy the motion equation $\ddot{x} + \omega^2 x =
\omega^2 (x_0 - a)$.
As clearly seen from (\ref{eq19}), particles will display
oscillations parallel to both the classical trajectory (which
oscillates around $x = a$ with frequency $\omega$) and other quantum
trajectories [i.e., given two particles, labeled as 1 and 2, their
respective trajectories will satisfy the property $x^{(2)}(t) -
x^{(1)}(t) = x_0^{(2)} - x_0^{(1)}$].
To understand how nonlocality operates here, note that the wave packet
does not spread with time but remains the same, with its center
following the path tracked by the corresponding classical trajectory.
This implies that the quantum motion is constrained.
Thus, since trajectories cannot cross, the only possibility for their
topology is to be the same as the classical one (different initial
positions will give different parallel trajectories).
That is, the nonlocality manifests in the distribution of initial
conditions, but not in the particular value of $\hbar$, since
$\omega t = \hbar t/2m\sigma_0^2$ can acquire any value.


\section{Conclusions}

As shown here, nonlocality is strongly related to the fact
that, at any time, the evolution of a (quantum) system strongly depends
on the full configuration of the {\it real} (coordinate) space, rather
than the features presented by the particular formulation used to
describe it.
In other words, the topology of the trajectories is strongly affected
by the information carried in the full
wave function (nonlocality could be thus described saying that {\it at
every time any particle has information about the whole system
configuration}, unless $\hbar \equiv 0$).
Therefore, the nonlocal behavior arises from having a complete
information encoded within the wave function, which is transmitted to
the (quantum) particles and indicates how they should evolve according
to certain rules as illustrated with the above two examples.
Due to the fact that the classical limit $\hbar \to 0$ is carried out
analytically, one can then observe at each time step the classical and
quantum behaviors.
Of course, as the system becomes more complex, the classical limit
is not carried out analytically and nonlocal effects are not easily
detected.
When mixed dynamics \cite{alberto} are proposed to solve problems with
high dimensionality, a detail analysis of the nonlocality should also
be carried out in order to give a complete interpretation of the results
in terms of trajectories because residual contributions, as those seen
in Eq.~(\ref{eq16}), can be masked by a fully classical analysis.

On the other hand, we would like to emphasize the insight that
Bohmian mechanics renders when applied to the study of physical
systems. In general, trajectory schemes can be of great help to
develop efficient computational algorithms to solve quatum problems.
However, at some point, one should also balance this efficiency with
the insight that they can provide.
Here we have seen that a WKB-BM scheme allows a nice comparison
between Bohmian and classical trajectories and, at the same time,
constitutes a method to tackle a quantum problem at different levels
of approximation depending on the (relative) value of $\hbar$.
In particular, this scheme is useful when trying to compare real
Bohmian trajectories with trajectories obtained from the WKB wave
function.
Generally, when using the WKB approximation, one only calculates the
first terms in the expansion \cite{sanz2,sanz21,sanz22},
$\bar{S}_0$ and $\bar{S}_1$,
which already contain nonlocality, as seen above.
Nevertheless, sometimes it could be necessary to go to higher orders
in semiclassical approximations as happens, for example, for deep
tunneling  where one needs more terms in order to account for the
longer tunneling times involved in the process \cite{pollak}.
Finally, we would like to emphasize that due to the large number of
works in WKB and WKB-like formalisms, the Bohmian analysis as
proposed here could be of great help to provide a deeper inside in
nonlocal effects.


\section*{Acknowledgements}

Support from the Spanish Ministry of Education and Science under the
project No. FIS2004-02461 is acknowledged.
A.S.\ Sanz would also like to thank the Spanish Ministry of Education
and Science for a ``Juan de la Cierva'' Contract.



\begin{thebibliography}
\eprint{}

\bibitem{paul1}
 P. Brumer, J. Gong, Phys. Rev. A 73 (2006) 052109.

\bibitem{grossman}
 F. Grossmann, Comments At. Mol. Phys. 34 (1999) 141.

\bibitem{miller}
 W.H. Miller, J. Phys. Chem. A 105 (2001) 2942.

\bibitem{eli}
 S. Zhang, E. Pollak, Phys. Rev. Lett. 91 (2003) 190201.

\bibitem{sanz4}
 R. Guantes, A.S. Sanz, J. Margalef-Roig, S. Miret-Art\'es,
 Surf. Sci. Rep. 53 (2004) 199.

\bibitem{paul21}
 C. Jaff\'e, P. Brumer, J. Phys. Chem. 88 (1984) 4829.

\bibitem{paul22}
 C. Jaff\'e, P. Brumer, J. Chem. Phys. 82 (1985) 2330.

\bibitem{paul31}
 J. Wilkie, P. Brumer, Phys. Rev. A 55 (1997) 27.

\bibitem{paul32}
 J. Wilkie, P. Brumer, Phys. Rev. A 55 (1997) 43.

\bibitem{alberto}
 E. Gindersperger, C. Meier, J. A. Beswick,
 J. Chem. Phys. 113 (2000) 9369.

\bibitem{leacock}
 R.A. Leacock, M.J. Padgett, Phys. Rev. D 28 (1983) 2491.

\bibitem{holland}
 P.R. Holland, The Quantum Theory of Motion,
 Cambridge University Press, Cambridge, 1993.

\bibitem{floyd1}
 E.R. Floyd, Phys. Rev. D 26 (1982) 1339.

\bibitem{floyd2}
 E.R. Floyd, Int. J. Mod. Phys. A 14 (1999) 1111.

\bibitem{faraggi1}
 A.E. Faraggi, M. Matone, Phys. Lett. A 249 (1998) 180.

\bibitem{faraggi2}
 A.E. Faraggi, M. Matone, Int. J. Mod. Phys. A 15 (2000) 1869.

\bibitem{john}
 M.V. John, Found. Phys. Lett. 15 (2002) 329.

\bibitem{wyatt1}
 C.-C. Chou, R.E. Wyatt, J. Chem. Phys. 125 (2006) 174103.

\bibitem{wyatt2}
  C.-C. Chou, R.E. Wyatt, Phys. Rev. E 74 (2006) 066702.

\bibitem{tannor}
 Y. Goldfarb, I. Degani, D.J. Tannor,
 J. Chem. Phys. 125 (2006) 231103.

\bibitem{sanz0}
 A.S. Sanz, S. Miret-Art\'es, J. Chem. Phys., submitted for
 publication.

\bibitem{sanz1}
 A.S. Sanz, F. Borondo, S. Miret-Art\'es,
 J. Phys.: Condens. Matter 14 (2002) 6109.

\bibitem{sanz10}
 A.S. Sanz, Ph.D. Thesis, Universidad Aut\'onoma de Madrid,
 Madrid, 2003.

\bibitem{sanz2}
 A.S. Sanz, F. Borondo, S. Miret-Art\'es, Europhys. Lett.
 55 (2001) 303.

\bibitem{sanz21}
 A.S. Sanz, F. Borondo, S. Miret-Art\'es,
 J. Chem. Phys. 120 (2004) 8794.

\bibitem{sanz22}
 A.S. Sanz, S. Miret-Art\'es,
 J. Chem. Phys. 122 (2005) 014702.

\bibitem{wyatt3}
 R.E. Wyatt, Quantum Dynamics with Trajectories,
 Springer, New York, 2005.

\bibitem{gottfried}
 K. Gottfried, Quantum Mechanics, W.A. Benjamin, New York, 1966,
 Vol.~1.

\bibitem{sanz7}
 A.S. Sanz, F. Borondo, S. Miret-Art\'es,
 Phys. Rev. B 61 (2000) 7743.

\bibitem{sanz8}
 A.S. Sanz, S. Miret-Art\'es, J. Chem. Phys. 126 (2007) 234106.

\bibitem{pollak}
 D.H. Zang, E. Pollak, Phys. Rev. Lett. 93 (2004) 140401.

\end{thebibliography}
\end{document}